# ChatGPT: ascertaining the self-evident. The use of AI in generating human knowledge

Ioannis D. Apostolopoulos [1,*], Mpesi Tzani [2], and Sokratis I. Aznaouridis [3]

1       Department of Medical Physics, School of Medicine, University of Patras, 26504 Rio, Greece; ece7216@upnet.gr

2       Department of Electrical and Computer Technology Engineering, University of Patras, 26504 Rio, Greece;

3       Department of Computer Engineering and Informatics, University of Patras, 26504 Rio, Greece

**Abstract:** The fundamental principles, potential applications, and ethical concerns of ChatGPT are analyzed and discussed in this study. Since ChatGPT emerged, it has gained a rapidly growing popularity, with more than 600 million users today. The development of ChatGPT was a significant mile-stone, as it demonstrated the potential of large-scale language models to generate natural language responses that are almost indistinguishable from those of a human. ChatGPT's operational principles, prospective applications, and ability to advance a range of human endeavours are discussed in the paper. However, much of the work discusses and poses moral and other problems that rely on the subject. To document the latter, we submitted 14 queries and captured the ChatGPT responses. ChatGPT appeared to be honest, self-knowledgeable, and careful with its answers. The authors come to the realization that since AI is already a part of society, the pervasiveness of the ChatGPT tool to the general public has once again brought to light concerns regarding AI in general. Still, they have moved from the domain of scientific community collective reflection at a conceptual level to everyday practice this time.

## 1. Introduction

ChatGPT is a large-scale artificial intelligence (AI) language model developed by OpenAI. The model was trained on a massive corpus of text data, including books, articles, and websites, using a technique known as deep learning. This approach enables the model to understand language structure and patterns and generate coherent responses to various queries [1].

The development of ChatGPT was a significant milestone, as it demonstrated the potential of large-scale language models to generate natural language responses that are almost indistinguishable from those of a human. Since its initial release, ChatGPT has been used for various applications, including chatbots, virtual assistants, and automated content generation [2].

While the development of ChatGPT and other large-scale language models has opened up exciting new possibilities for AI, it has also raised concerns about the potential impact of the technology on society [3]. For example, some researchers have raised concerns about the potential for these models to perpetuate bias and discrimination, as they are trained on data that may reflect historical or cultural biases [4–6].

To address these concerns, OpenAI and other organizations have developed ethical guidelines for developing and deploying AI, including language models [7]. These guidelines emphasize the importance of transparency, accountability, and responsible use of AI and encourage researchers and developers to consider the potential impact of their work on society as a whole.

The paper presents how ChatGPT works, its possible applications, and the potential to contribute to various fields of human activity. Much of the work, however, presents and comments on ethical and other dilemmas that arise depending on the area. The authors conclude that the diffusion of the ChatGPT tool to the general public brought back to light concerns regarding artificial intelligence in general, but this time passing from the sphere of collective reflection of scientists at a theoretical level to everyday practice since AI is already part of society.

## 1.1. Related Work

The scientific community has rapidly responded to the emergence of ChatGPT. Most published papers have paid attention to potential applications and ethical concerns.

In [8], the author presented some concerns about the ChatGPT use regarding education and academic writing. It is highlighted that academic reviewers only caught 63% of AI-generated abstracts. Moreover, the author stated that "the implications for education may push academics to rethink their courses innovatively and give assignments that AI doesn't easily solve".

Rudolf et al. [9] presented the story of OpenAI and experimented with the software tool for educational purposes. In their paper, the authors briefly introduced the potential and limitations of ChatGPT and discussed the future of learning, teaching and assessment in higher education in the context of AI chatbots such as ChatGPT. The paper is ideally suited for teachers and teaching-policy planners as it provides a good understanding of how such tools work and discusses the potential use and misuse of chatbots in various aspects.

Liebrenz [10] stated that ChatGPT might democratize the dissemination of knowledge since it circumvents language barriers. The authors point out potentially inaccurate outputs. The authors presented a conversation with ChatGPT, wherein a human requested that ChatGPT generate a response regarding what the Lancet editing team should do with AI-generated texts. Surprisingly, ChatGPT showed great self-knowledge, exposing its flaws and current ethical concerns.

In [2], the authors evaluated the performance of ChatGPT on the United States Medical Licensing Exam (USMLE). It performed near the passing threshold for every part of the exam without any prior preparation or specialized focus. The authors concluded that such language models have the potential to assist with medical education and might prove helpful to clinical decision-making.

In [7], the authors presented issues for discussion about language models. For example, the kind of research tasks to be outsourced to language models, the academic characteristics that remain unique and essential to researchers, the potential of human verification when using such tools, the way to ensure that equity in research is promoted, and legal implications are among the raised issues.

## 2. Materials and Methods

### 2.1. Generative Pretrained Transformer

GPT stands for Generative pretrained Transformer [11–13], an artificial neural network that uses deep learning to analyze and generate natural language. GPT models are pre-trained on large amounts of text data and can generate human-like text in response to prompts [1]. GPT models have been used in various applications, including natural language processing, language translation, text completion, and chatbots [6]. ChatGPT is a specific instance of the GPT architecture designed to generate human-like responses to natural language prompts in a casual setting.

The architecture of GPT is based on a deep learning [14] model known as the Transformer. The Transformer model is a type of neural network that uses self-attention mechanisms to process data sequences, such as text [15].

The core component of the Transformer model allows it to focus on different parts of the input sequence at different timesteps. The self-attention mechanism [16,17] assigns weights to each element in the sequence based on its relevance to the other elements. The model can capture long-range dependencies and relationships between input components by attending to different parts of the input sequence.

Generative AI is artificial intelligence capable of creating new and original content rather than simply classifying or recognizing existing data. In the case of GPT, the model is trained on a large corpus of text data and can generate new text in response to a given prompt. The model uses a probabilistic approach to generate text, which generates a probability distribution over the following words or tokens in the sequence. The most likely next token is selected based on this distribution, and the process is repeated to generate longer text sequences.

One of the critical advantages of GPT is its ability to be pre-trained on large amounts of text data before being fine-tuned for specific natural language processing tasks. This pre-training process allows the model to learn general patterns and relationships in language. However, GPT also has some limitations, such as the potential for generating biased or misleading content based on the data it was trained on [5].

### 2.2. The ChatGPT tool: history, methods, and applications

The origins of ChatGPT can be traced back to the early days of OpenAI, founded in 2015 by a group of researchers and entrepreneurs. The company aimed to advance AI to benefit humanity while mitigating potential risks associated with the technology.

The key to ChatGPT's success is its ability to learn from large volumes of text data using a process known as unsupervised learning. Unlike traditional machine learning approaches, which require labelled data to train models, unsupervised learning enables models like ChatGPT to learn from raw, unstructured data by identifying patterns and relationships between words and phrases. Figure 1 displays some of the statistics on ChatGPT.

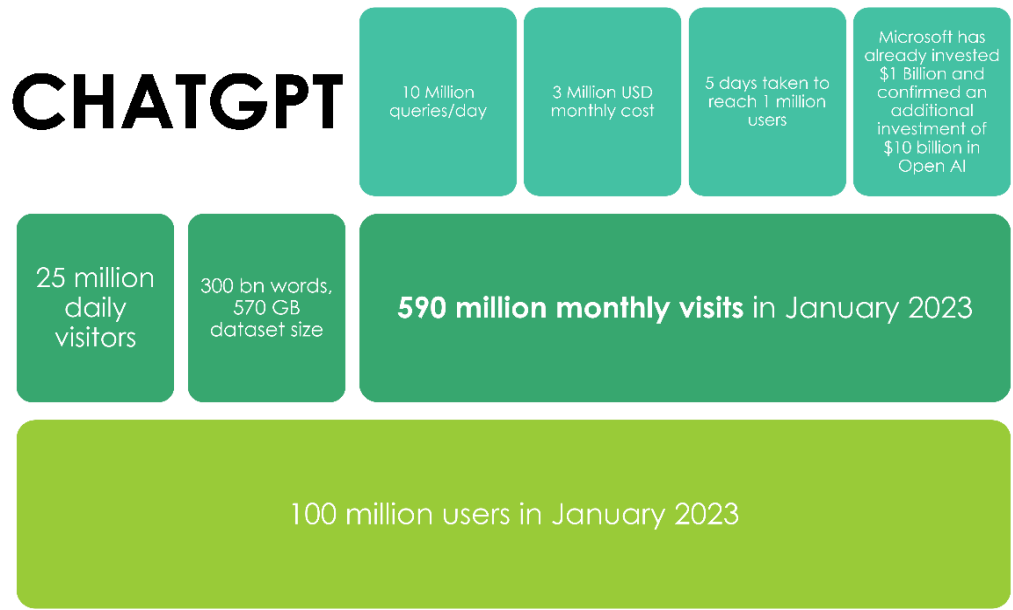

Figure 1. CHATGPT statistics at the end of January 2023. Source: nerdynav.com

ChatGPT has a wide range of potential applications across various industries, from customer service to education, entertainment, and more. Some of the most promising applications of ChatGPT are presented as follows:

Customer Service and Support: ChatGPT can be used to develop virtual customer service agents, chatbots, and virtual assistants that can respond to customer inquiries and resolve issues in real time. For example, companies like Capital One and Bank of America use chatbots to assist customers with their banking needs.

Education: ChatGPT can be used to develop intelligent tutoring systems to help students with their homework and test preparation. For example, an AI-powered writing assistant powered by ChatGPT can provide students with feedback on their essays, help them improve their writing skills, and suggest relevant resources for further study.

Healthcare: ChatGPT can be used to develop virtual health assistants to provide patients with personalized health advice and recommendations. For example, the AI-powered chatbot Woebot uses ChatGPT to provide users with cognitive behavioural therapy for mental health issues like anxiety and depression.

Entertainment: ChatGPT can be used to develop chatbots and virtual assistants that can provide personalized recommendations for movies, TV shows, and other forms of entertainment. For example, the chatbot Mitsuku uses ChatGPT to chat with users and provide them with movie recommendations based on their preferences.

E-commerce: ChatGPT can be used to develop virtual shopping assistants to provide customers with personalized product recommendations and assist with purchasing. For example, H&M uses a chatbot powered by ChatGPT to help customers find the right outfit for their needs.

Journalism and Content Creation: ChatGPT can be used to develop automated content creation systems that can write news articles, blog posts, and other forms of content. For example, The Washington Post has created an AI-powered tool called Heliograf that uses ChatGPT to write news articles based on data and other sources.

Overall, the potential applications of ChatGPT are vast and varied, and the technology is likely to significantly impact various industries in the coming years. As technology continues to evolve and improve, it will be interesting to see how organizations and individuals leverage its capabilities to achieve their goals and improve their operations.

### 2.3. Experiment Design

The authors created an account in OpenAI and accessed the free version of ChatGPT. The used version was ChatGPT Feb 13. The study involves 14 queries and 14 responses. The latter are presented in the exact way they were formed by the tool, with-out performing any grammatical and error corrections. The first seven queries are preliminary so we can get to know ChatGPT better and chat with it. The remaining queries are technical and challenging, because they request information on recent AI algorithms and XAI.

The AI model is assessed and the conclusions are analytically discussed based on the responses.

## 3. Results

The authors point out that the text after the present statement is entirely generated by ChatGPT following specific queries. These queries were posed by the authors.

### 3.1. An interview with ChatGPT

*3.1.1. Query 1: Who are you?"*

I am an AI language model created by OpenAI. I have been trained on a large corpus of text and I am able to generate responses to questions and provide information on a wide range of topics. My goal is to assist users and provide them with accurate and helpful information.

*3.1.2. "Query 2: Nice to meet you. I am John."*

Nice to meet you, John! How can I help you today?

### 3.1.3. "Query 3: ChatGPT who made you?"

I was created by OpenAI, an artificial intelligence research laboratory consisting of the for-profit technology company, OpenAI LP, and its parent company, the non-profit OpenAI Inc. OpenAI was founded in 2015 with the goal of promoting and developing friendly AI that benefits humanity. OpenAI aims to advance digital intelligence in a responsible and safe way, and it has become one of the leading organizations in the field of AI research and development.

### 3.1.4. "Query 4: What do you know about me?"

As an AI language model, I don't have access to personal information about individuals unless it has been explicitly shared with me. I can only respond based on the information provided by the user in the current conversation.

### 3.1.5. "Query 5: Do you believe that you give the correct answers from a scientific perspective?"

As an AI language model, I don't have beliefs or personal opinions. My goal is to provide accurate and helpful information based on the data that I have been trained on. The information I provide is based on publicly available sources and the most up-to-date knowledge available at my training data's knowledge cut-off point, which was 2021. It is important to note that not all information found on the internet is accurate, and I may not always be able to provide the most up-to-date information or the most accurate answers. When in doubt, it is always best to consult multiple credible sources and consult with subject matter experts.

### 3.1.6. "Query 6: But the internet does not contain only trusted sources. How do you detect the most reliable ones?"

As an AI language model, I don't have the ability to independently evaluate the reliability or credibility of a source. I was trained on a diverse range of text data from the internet, which includes both reliable and unreliable sources. My training data includes information from sources such as news articles, scientific journals, and encyclopedias, which tend to be more reliable. However, it also includes information from sources such as personal blogs, forums, and user-generated content, which may not be as reliable.

When providing information, my goal is to present the most relevant and accurate information based on the context of the question. I try to provide a balanced perspective and include information from multiple sources whenever possible. However, it is always a good idea to verify information with multiple credible sources and to consult subject matter experts when in doubt.

### 3.1.7. "Query 7: Can I trust you for writing my essay?"

As an AI language model, I can assist with generating text and providing information on a wide range of topics. However, it is not recommended to solely rely on my responses for writing academic papers or assignments.

While I can provide information and generate text that may be useful as a starting point or reference, it is important to carefully review and fact-check the information I provide. Additionally, academic writing often requires original thoughts, analysis, and arguments, which are best developed through independent research and critical thinking.

In academic settings, it is also important to properly cite all sources used in writing, including any information provided by me. Plagiarism, which includes submitting someone else's work as one's own, is taken very seriously in academic institutions and can have serious consequences.

In summary, while I can be a useful resource for generating text and providing information, it is important to independently verify the information I provide and to engage in original research and analysis when writing academic papers or assignments.

## 3.2. ChatGPT in action

### 3.2.1. "Query 8: Describe artificial intelligence brief history in three paragraphs"

The history of Artificial Intelligence (AI) dates back to the 1950s when British computer scientist and mathematician, Alan Turing, proposed the concept of machine intelligence and its potential to perform tasks normally done by humans. In 1956, a group of computer scientists and cognitive psychologists gathered at Dartmouth College, New Hampshire, to discuss the possibility of building "thinking machines." This event is often considered the birth of AI as a field of study.

In the following decades, AI research saw significant advances, including the development of expert systems, which could perform specific tasks with human-level expertise, and the creation of the first AI programming language, LISP. However, AI faced setbacks in the 1980s when funding was cut and the field's initial promises failed to materialize. It wasn't until the late 1990s and early 2000s that AI began to gain momentum again, thanks to breakthroughs in machine learning and access to large amounts of data and computing power

Today, AI is a rapidly growing field with a wide range of applications, from consumer products like smart speakers and virtual assistants to complex systems for healthcare, finance, and transportation. As AI continues to evolve and become more sophisticated, it raises important ethical and societal questions about the role of machines in our lives and the potential impact on jobs and the economy. Nevertheless, AI has the potential to greatly benefit society, and it remains an exciting and rapidly advancing area of research and development.

### 3.2.2. *"Query 9: Describe what is machine learning and deep learning in two paragraphs"*

Machine Learning (ML) is a subfield of AI that focuses on enabling computers to learn from data without being explicitly programmed to perform specific tasks. The goal of ML is to develop algorithms and models that can automatically improve their performance as they are exposed to more data. There are several types of ML, including supervised learning, unsupervised learning, and reinforcement learning. In supervised learning, the algorithms are trained on labeled data, meaning the data is categorized into specific classes, and the goal is to predict the class of new, unseen data. In unsupervised learning, the algorithms are trained on unlabeled data, and the goal is to find patterns or structure in the data. In reinforcement learning, the algorithms learn through trial and error by receiving rewards or penalties for certain actions.

Deep Learning (DL) is a subfield of ML that is based on artificial neural networks, which are inspired by the structure and function of the human brain. Deep Learning algorithms consist of multiple layers of interconnected nodes that process and analyze data. The layers in a deep learning model are often referred to as "hidden layers" because they are not directly connected to the inputs or outputs of the model. The strength of Deep Learning lies in its ability to automatically extract complex features from large amounts of data and improve its accuracy over time. This has led to remarkable breakthroughs in fields such as computer vision, natural language processing, and speech recognition, and has paved the way for exciting new applications in areas such as autonomous vehicles and personalized medicine.

### 3.2.3. *"Query 10: Describe the convolutional neural networks from a mathematical perspective in detail"*

Convolutional Neural Networks (ConvNets or CNNs) are a type of neural network that are particularly well suited for image and video recognition tasks. At their core, ConvNets use a mathematical operation called convolution to analyze the relationship between elements in an image and identify features that are important for classification.

A convolution operation involves a filter, which is a small matrix of values, and a feature map, which is a representation of the input image or a previous layer in the network. The filter is slid across the feature map, and at each position, the values in the filter are multiplied by the corresponding values in the feature map and then summed to produce a new value. This process is repeated for every position of the filter, and the result is a new feature map, known as a convolved feature map, that summarizes the information in the original feature map.

The convolved feature maps are then processed by other layers in the network, including pooling layers, which down-sample the feature maps by taking the maximum or average value in a region of the feature map, and fully connected layers, which produce the final prediction. The weights in the network are learned through a process called backpropagation, where the error in the prediction is propagated backwards through the network and used to update the weights in an attempt to minimize the prediction error.

The combination of convolution and pooling operations in ConvNets allows for a hierarchical analysis of the input data, where low-level features such as edges and corners are combined to form higher-level features such as object parts, objects, and scenes. ConvNets can also be trained end-to-end, meaning that the weights in the network are learned directly from the input data, which makes them well suited for tasks where the relationship between the inputs and outputs is complex and difficult to specify manual-ly.

"Query 11: Give me some equations on the above query"

Sure, here are some of the main equations used in Convolutional Neural Networks (ConvNets) to perform the convolution and pooling operations:

Convolution:
Let the input feature map be represented by X and the filter be represented by W. The convolution operation can be defined as follows:

$$Z = W * X + b,$$

where Z is the resulting convolved feature map, * represents the element-wise multiplication, and b is a bias term. The convolution operation is performed for each position of the filter, and the resulting values are stored in Z.

Pooling:

Pooling is used to down-sample the feature maps and reduce the spatial dimensions of the data. Commonly used pooling operations include max pooling and average pooling. Max pooling selects the maximum value in a region of the feature map, while average pooling takes the average of the values in a region.

Max pooling:

Let X be the input feature map and S be the pooling size. The max pooling operation can be defined as follows:

$$Z = \max(X[i:i+S, j:j+S]),$$

where Z is the resulting down-sampled feature map, i and j are the row and column indices of the feature map, and max() is the maximum operator.

Average pooling:

The average pooling operation can be defined similarly to max pooling, except that the average of the values in a region is taken instead of the maximum value.

$$Z = \mathrm{mean}(X[i:i+S, j:j+S]),$$

where mean() is the mean operator.

These equations give a basic overview of the convolution and pooling operations used in ConvNets. However, in practice, ConvNets may use multiple filters and multiple pooling layers, and the exact implementation can vary depending on the task and architecture being used.

### 3.2.4. "Query 12: What is explainable artificial intelligence? "

Explainable Artificial Intelligence (XAI) refers to the development of AI systems that are able to provide clear and interpretable explanations of their decision-making processes. XAI is an important aspect of AI development because it helps to increase the transparency and accountability of AI systems, which can improve trust and confidence in their results.

The goal of XAI is to provide insights into the workings of AI systems, making it possible to understand why the system arrived at a particular decision, and to identify any biases or errors in its decision-making process. XAI techniques can include visualizations of the inputs and intermediate representations of the data in the AI system, as well as the use of methods such as sensitivity analysis and feature importance to under-stand the impact of individual inputs on the system's outputs.

XAI is particularly important in high-stakes applications such as healthcare, finance, and criminal justice, where the consequences of incorrect decisions can be significant. In these cases, having a clear understanding of the reasoning behind an AI decision can be essential for ensuring that the results are accurate and fair. Additionally, XAI can help to address concerns about the ethics of AI and to ensure that AI systems are developed in a responsible and sustainable manner.

### 3.2.5. *"Query 13: I want to know more. please inform me about major methods and how they operate"*

There are several major methods used in Explainable Artificial Intelligence (XAI) to provide insights into the workings of AI systems. Some of these methods include:

Visualizations: Visualizations of the inputs and intermediate representations of the data can provide a way to understand what the AI system is looking at and how it is processing the data. For example, saliency maps can be used to highlight the areas of an image that are most important to the system's predictions. Activation maps can also be used to show which neurons in a neural network are most active for a particular input.

Feature Importance: Feature importance methods, also known as feature attribution methods, are used to understand the impact of individual inputs on the system's out-puts. These methods can include techniques such as feature permutation, where the values of individual inputs are perturbed, and the resulting change in the output is used to determine the importance of that input.

Sensitivity Analysis: Sensitivity analysis methods are used to understand how changes to the inputs of the system affect its outputs. For example, local interpretable model-agnostic explanations (LIME) generate explanations for a specific prediction by fitting a simple model to the data around that prediction and then analyzing the importance of individual inputs in that model.

Model Inspection: Model inspection involves directly examining the structure and parameters of the AI system. For example, decision trees can provide a visual representation of the logic used by the system to make predictions, and rule-based systems can provide a clear set of rules that can be used to understand the system's behavior.

These methods provide different perspectives on the workings of AI systems, and they can be used in combination to provide a comprehensive understanding of the system's behavior. The choice of which method to use depends on the specific requirements of the task and the nature of the AI system being used. Overall, the goal of XAI is to make AI systems more transparent and interpretable, and to increase the accountability and trust in their results.

### 3.2.6. *"Query 14: Tell me about your flaws with examples. Some say you may be biased, give me 3 examples. Some say you are introducing discrimination. How?"*

As an AI language model, I am not capable of having biases or prejudices because I do not have personal beliefs, opinions, or feelings. However, I can inherit biases and prejudices that exist in the training data that I was trained on, or that may be intro-duced by users who interact with me. Here are some examples of how these issues can arise:

Biased training data: AI language models like me are trained on large datasets of text, which can include biased or prejudiced language. For example, if the training data contains a disproportionate amount of text written by white male authors, the model may be more likely to generate text that reflects their perspectives and experiences, and may not accurately represent the perspectives of people from other backgrounds.

User-generated content: Another way that biases can be introduced is through the text that users input when interacting with me. If users input language that contains biases or prejudices, such as using discriminatory language or making stereotypical assumptions, I may generate text that reflects these biases. However, it's important to note that I don't create content on my own; I only generate responses based on the input I receive from users.

Algorithmic bias: Finally, AI language models can also exhibit algorithmic biases that are inherent in the way they are designed to operate. For example, if the model is optimized for certain performance metrics, it may prioritize certain types of responses over others, which can lead to biased or unrepresentative results.

It's important to note that these issues are not unique to me as an AI language model; they are common challenges that arise in many areas of machine learning and AI development. However, they are important to recognize and address in order to ensure that AI is developed and used in an ethical and responsible manner.

## 4. Discussion

ChatGPT was asked to respond to 14 queries. These are summarized in **Table 1**. The model's responses are suitable for reaching some conclusions. Highlights are also presented in **Table 1**.

**Table 1**. Queries and response highlights

| No | Query | Highlights |
|---|---|---|
| 1 | Who are you? | The response is honest, and the model expresses its goals |
| 2 | Nice to meet you. I am John. | - |
| 3 | ChatGPT who made you? | The model knows its creators |
| 4 | What do you know about me? | The model informs the user that it does not have access to personal data |
| 5 | Do you believe that you give the correct answers from a scientific perspective? | The model knows its sources and states that it has no personal beliefs |
| 6 | But the internet does not contain only trusted sources. How do you detect the most reliable ones? | The model does not take responsibility for misleading information |
| 7 | Can I trust you for writing my essay? | The model recommends its careful use |
| 8 | Describe artificial intelligence brief history in three paragraphs | The model deviates from presenting the history and mentions ethical concerns, which are irrelevant to the query |
| 9 | Describe what is machine learning and deep learning in two paragraphs | The model responds successfully, though it fails to mention the concept of semi-supervised learning. The model does not provide any references |
| 10 | Describe the convolutional neural networks from a mathematical perspective in detail | The model confuses the mathematical operation of convolution with convolution in CNNs |
| 11 | Give me some equations on the above query | The model selected the operations on convolution, max and average pooling and excluded the activation function and min pooling, which are equally essential. The presented equations are correct. |
| 12 | What is explainable artificial intelligence? | The response is correct |

| 13 | I want to know more. please inform me about major methods and how they operate | The response is correct |
| 14 | Tell me about your flaws with examples. Some say you may be biased, give me 3 examples. Some say you are introducing discrimination. How? | The model has self-knowledge. It also realizes that it shares the limitations of roughly every AI model |

We could summarize the conclusions based on the model's responses as follows:

ChatGPT knows its creators and can express its goals, as demonstrated by the responses in queries one and three.

The model does not use personal data for generating the responses, as proved by the answer we received from query four.

The model knows its sources and states that it has no personal beliefs (query five).

The model does not take responsibility for misleading information and warns the user about it (query six).

As demonstrated in query seven's response, the model knows its limits and instructs the user to exploit its capabilities carefully.

The model is not precise in describing the history of AI. It deviates from presenting the history and mentions ethical concerns irrelevant to the query (query 8). In addition, it responds successfully when asked to describe the notions of ML and DL. However, it fails to mention the concept of semi-supervised learning. Moreover, it does not provide any references. In query 11, ChatGPT selected the operations on convolution, max and average pooling, excluding the activation function and min pooling, which are equally essential. The presented equations are correct. ChatGPT gave correct details about XAI and its primary methods (queries 12 and 13).

The model has self-knowledge. It also realizes that it shares the limitations of roughly every AI model. Its response to query 14 is remarkable and shows excellent potential from the AI perspective.

One tremendous setback of most AI models, including ChatGPT, is their black-box nature [18–20]. Users are not aware of any hidden process and can't perceive how the model decided to generate the particular response. Moreover, the users are not informed about how ChatGPT encapsulates informative content from previous conversations and how it constitutes a user profile to deliver more personalized responses.

## 4.1. Ethical Concerns

There are worries about the responsibility and ownership of content produced through AI writing tools like ChatGPT. Even though such tools can lower instances of plagiarism, human editing is still necessary to completely eliminate it. Furthermore, using AI-generated letters of recommendation or personal statements may spark questions about their legitimacy.

It's crucial to ensure that using AI-generated text for commercial purposes does not violate any existing copyrights when using AI-generated text. As AI-generated text is based on pre-existing data, there may be a risk of repetition and a lack of imagination in the text. Using AI-powered text generation can result in a loss of creative and unique content that human authors usually provide. Additionally, the ease of using AI to generate text can also lead to decreased student engagement and motivation in educational institutions.

One worry is that AI-generated text may have inaccuracies or mistakes. The current ChatGPT does not have a way to assess the correctness of its output. There is the worry that AI-generated text may perpetuate or increase existing biases, as the models are trained on data sets that may contain biases. It is crucial for the use of AI in writing to be clearly identified and transparent so that readers are aware of the source of the text.

One ethical concern surrounding the use of ChatGPT is the potential for the technology to be used for social engineering or deception. ChatGPT could be used to create convincing impersonations of individuals or organizations, which could spread false information, manipulate public opinion, or carry out fraudulent activities. Additionally, the use of ChatGPT in customer service chatbots and other applications could potentially displace human workers and contribute to job loss and economic inequality.

Another ethical concern is the potential for ChatGPT to perpetuate bias and discrimination in language. Because ChatGPT is pre-trained on large amounts of text data, it may learn and reinforce patterns of prejudice and discrimination that exist in the data. It is essential to carefully monitor and evaluate the output of ChatGPT to ensure that it is not perpetuating harmful biases or stereotypes. Various ethical concerns are illustrated in Figure 2.

We present particular concerns in various fields:

Mental health: One ethical concern is the potential for ChatGPT to replace human therapists, which may not be able to provide the same level of care and support. Additionally, there may be privacy concerns related to collecting and storing sensitive personal information.

Education: There are concerns about the potential for ChatGPT to facilitate cheating and the potential for bias or misinformation in the information provided. Additionally, there may be concerns about student data privacy and the collection and use of student information.

Customer service: There are concerns about the potential for bias in the responses provided and the potential for ChatGPT to manipulate customers or collect personal information without their consent. Additionally, human customer service representatives may be concerned about job displacement.

Legal: There are concerns about the potential for ChatGPT to provide inaccurate information or advice and the potential for bias in the information provided. Additionally, there may be concerns about client confidentiality and protecting sensitive legal information.

Medicine: Ethical concerns about using chatbots in medical contexts include potential privacy violations of patient data, confidentiality issues, and the potential for a system to miss important cues and provide incorrect guidance due to a lack of medical expertise. Additionally, doctors may be reluctant to provide a less personal level of care when using a chatbot, leading to a lack of trust and communication between the doctor and the patient. Additionally, since chatbot technologies are relatively new, there is no clear consensus on ethically structuring these technologies in healthcare.

Policy-making: Ethics are important when using advanced AI technology like ChatGPT for decision-making. Human experts should always supervise ChatGPT to ensure it can be held accountable for its decisions.

Entertainment: Ethical concerns about using AI in entertainment primarily revolve around the potential for AI to be used to produce content that is deemed inappropriate or offensive. AI-written content can be subject to bias and harm various minority groups. Additionally, there is a risk of AI-written content being mistaken for genuine human writing, which could lead to potential embarrassment or legal repercussions. Furthermore, lacking human oversight may lead to flawed content that may fail to meet the required accuracy, quality, and context standards. For example, an AI-written article on a controversial topic might be produced that is inaccurate, factually incorrect, or does not fit the desired narrative. It is important to ensure that AI-written content is regularly monitored and assessed for accuracy, quality, and appropriateness.

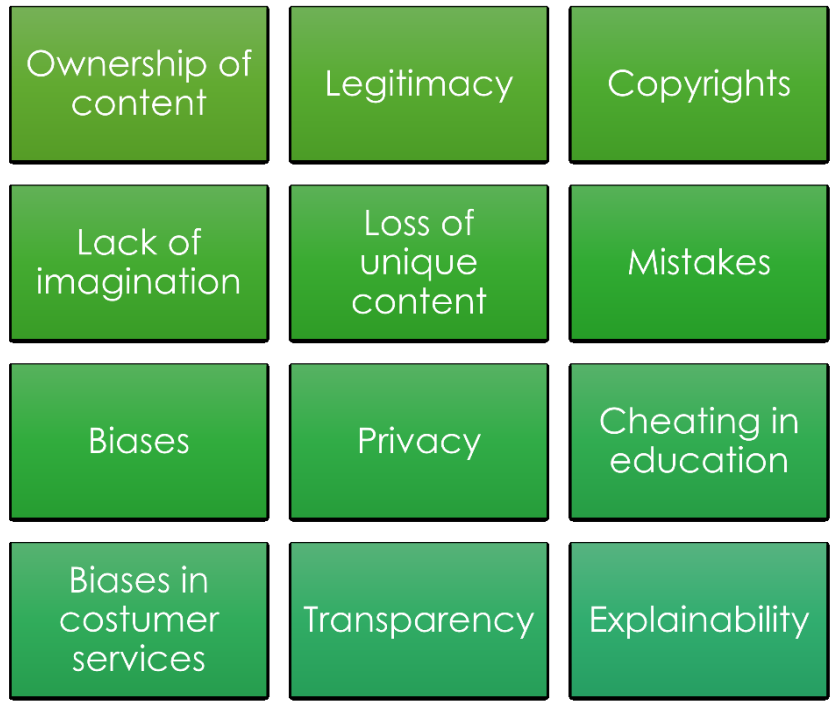

Figure 2. ChatGPT concerns

## 5. Conclusions

The paper details ChatGPT's operation, potential applications, and the opportunities it may provide. Much of the work, however, addresses ethical and other issues that could arise in various areas of human activity. The study challenged ChatGPT with 14 queries and captured and analyzed responses. The study demonstrated with examples that ChatGPT has self-knowledge regarding its biases, the possibility of generating in-correct responses, and its training data. It is also informative to the reader and tries to avoid misleading information. It is also careful in the responses and tries to prevent provocative queries that may generate biased responses.

ChatGPT shares the same limitations as roughly any ML model. It is a complete black box. The reader does not have a single piece of evidence of its reasoning, how it memorizes and uses the user's behaviour and past questions, and how it reaches a decision. The use of ChatGPT comes with several ethical concerns, as most AI frameworks.

ChatGPT has many potential applications, including customer service chatbots, virtual assistants, and personalized conversational agents. The authors argue that ChatGPT's diffusion to the general public brought social unease to mind over AI in the past. However, now it happens at a more concrete level, as AI is already being utilized in everyday life.

## Statements and Declarations

### Competing Interests

The authors declare that they have no conflict of interest. The authors have no relevant financial or non-financial interests to disclose.

### Funding

The research received no external funding